\documentclass[prc,showpacs,floatfix,preprint]{revtex4-1}
\usepackage{amsmath}
\usepackage{epsfig}
\usepackage{color}
\usepackage{textcomp}

\newcommand{\be}{\begin{equation}}
\newcommand{\ee}{\end{equation}}

\newcommand{\ber}{\begin{eqnarray}}
\newcommand{\eer}{\end{eqnarray}}

\begin{document}

\title{Determination of $S$-factors with the LIT method}

\author{ Winfried Leidemann$^{1,2}$,  Sergio Deflorian$^{1,2}$ and Victor D. Efros$^{3,4}$ 
  }

\affiliation{
  $^{1}$Dipartimento di Fisica, Universit\`a di Trento, I-38123 Trento, Italy \\
  $^{2}$Istituto Nazionale di Fisica Nucleare, TIFPA,
  I-38123 Trento, Italy \\
  $^{3}$ National Research Centre "Kurchatov Institute", 123182 Moscow, Russia\\
  $^{4}$ National Research Nuclear University MEPhI(Moscow Engineering Physics Institute),
  Moscow, Russia   
}

\begin{abstract}
The precise determination of astrophysical $S$-factors is essential for a
detailed understanding of the nucleosynthesis in its various facets.
It is discussed how the Lorentz integral transform (LIT) method can be 
applied for such a determination.
The astrophysical $S$-factor for the proton-deuteron radiative capture is considered as test case. The importance of a specific
many-body basis used for the LIT equation solution is pointed
out. The excellent results of the test are discussed.   
\end{abstract}

\maketitle

\section{Introduction}
\label{intro}
Cross sections of astrophysical reactions become generally quite small in presence of 
a Coulomb barrier between the reactants. Thus it can become very difficult, 
or almost impossible, to determine the corresponding $S$-factors experimentally.
A possible resort is a theoretical determination, but then
one has to deal with a continuum-state problem, which, however, can be avoided
using the Lorentz integral transform  \cite{EfL07}. In fact applying the LIT 
one is confronted only with a bound-state like problem. 

To check the 
precision of such a LIT calculation we consider the $S$-factor of the reaction $^2$H$(p,\gamma)^3$He,
\begin{equation}
\label{S-fac}
 S_{12}(E) = \sigma_{\rm cap}(E) \, E \,\exp(2\pi\eta) \,,
\end{equation}
as test case. In the equation above
$\sigma_{\rm cap}$ is the reaction cross section and
$E$ is the relative energy of the deuteron-proton pair, whereas $\exp(2\pi\eta)$ denotes 
the well known Gamow factor which takes into account the effect of the Coulomb barrier.
We determine $\sigma_{\rm cap}$ by using time reversal invariance  
thus calculating the cross section $\sigma_\gamma$ of the inverse reaction 
$^3$He$(\gamma,p)^2$H.

In this work it is not intended to give realistic results for $S_{12}$.
As already mentioned we rather perform a test whether $S$-factors can be reliably
calculated with the LIT method. Therefore we compute the $^3$He photodisintegration 
cross section just in unretarded dipole approximation and use a simple
$NN$ interaction, namely the central MT-I/III potential~\cite{MaT69}. In order to check our
LIT results we also make a conventional calculation with explicit continuum wave 
functions which are calculated using correlated hyperspherical harmonics (for details
of this calculation see \cite{DeE16}).

\section{Formalism}

For a given photon energy $E_\gamma$ the unretarded dipole cross section 
\begin{equation}
\label{xsec}
\sigma_\gamma(E_\gamma) = 4 \pi^2 \alpha E_\gamma R(E_\gamma) 
\end{equation}
is determined via the calculation of the dipole response function
\begin{equation}
\label{response}
R(E_\gamma) = \int df |\langle f| D_z | 0\rangle|^2 \delta(E_f - E_0 - E_\gamma) \,, 
\end{equation}
where $|0 \rangle$ and $|f\rangle$ denote the 
$^3$He ground state and the deuteron-proton continuum state, respectively, 
$E_0$ and $E_f$ are the corresponding eigenenergies, whereas
$D_z$ is the third component of the nuclear dipole operator.
The LIT of the response function $R(E_\gamma)$ is defined by
\begin{equation}
\label{LIT}
L(\sigma=\sigma_R + i\sigma_I) = \int_{E_{thr}}^\infty dE_\gamma \, {\frac {R(E_\gamma)}{(E_\gamma-\sigma_R)^2 + \sigma_I^2}}\,.
\end{equation}
It is important to realize that the LIT is an integral transform with a controlled 
resolution. This is due to the variable width of 2$\sigma_I$ of the Lorentzian kernel,
which, in principle, can be adjusted to resolve the detailed structure of $R(E_\gamma)$. 
On the other hand choosing an increased resolution by a reduced $\sigma_I$ requires in 
general an increase of the precision of the calculation.

With a Hamiltonian $H$ of the three-nucleon system the LIT   
of Eq.~(\ref{LIT}) is calculated via the equation
\begin{equation} 
\label{eqLIT}
(H-E_0-\sigma) \, |\tilde\Psi(\sigma)\rangle =  D_z | 0\rangle \,.
\end{equation}
Since the solution $\tilde\Psi(\sigma)$ is localized, it can be obtained using 
bound-state methods. The transform is then given by
\begin{equation}
\label{LIT1}
L(\sigma) = \langle \tilde\Psi(\sigma) | \tilde\Psi(\sigma) \rangle \,.
\end{equation}
Finally, the response $R(E_\gamma)$ is obtained from the inversion of
the calculated $L(\sigma)$ (for details see \cite{EfL07}).

The LIT approach has been used for quite a few calculations of electromagnetic
nuclear responses. In most of these cases a hyperspherical harmonics (HH) basis
has been employed for the expansions of nuclear ground-state wave functions
and LIT states $\tilde\Psi$. However, it has been realized in \cite{Lei15} that
it can be more advantageous to use a different basis in a kinematical region
where the nucleus can be disintegrated exclusively into two fragments. For
a better explanation of this fact let us rewrite the LIT of Eq.~(\ref{LIT1})
in a different form:
\begin{equation}
\label{LIT_En}
 L(\sigma) = \sum_{n=1}^N {\frac { |\langle \phi_n| D_z | 0 \rangle |^2}
    {(\sigma_R-(E_n-E_0))^2 + \sigma_I^2}} \,. 
\end{equation}
In the equation above $\phi_n$ and $E_n$ are the eigenfunctions and eigenvalues 
resulting from the diagonalization of the Hamiltonian on a given many-body basis with 
dimension $N$. Let us furthermore define 
$\Delta E_n = E_{n+1} - E_n$, which is a measure for the density of LIT states
in a given energy region. Now, using an HH basis, it is very difficult to 
systematically decrease the $\Delta E_n$ below the three-body breakup threshold. 
This can lead to problems at lower energies if only two-body breakup channels are
open. In fact if one wants to resolve with the LIT approach a 
cross section structure of width $\Gamma$ at an energy $E_x$ one needs that the 
$\Delta E_n$ around $E_x$ should not be larger than $\Gamma$. With a basis different from
the HH basis it is however possible to systematically increase the density of LIT 
states in the two-body breakup region. It consists in an HH basis for the (A-1) nucleon 
system with an additional basis for the wave function of the A-th nucleon with respect 
to the center-of-mass of the (A-1) nucleon system. In this way it became possible to 
determine the width of the $^4$He isoscalar monopole resonance to 180(70) keV, 
which compares quite well with the experimental result from inelastic electron 
scattering \cite{Lei15}. 

For the present aim, the determination of the $S$-factor $S_{12}$, one is not confronted
with a resonant behaviour, but with a cross section that has as main feature an 
exponential increase at threshold. Since we intend to determine details of the cross 
section with great precision even beyond the exponential increase it is desirable 
to have a dense grid of $\sigma_R$-values in the low-energy region.

\section{Results and discussion}

According to what we said at the end of the previous section we should use for the 
three-body LIT calculation an HH basis for the (3-1)-system. Of course for such a
two-body system the HH basis reduces to a basis with normal spherical harmonics.
We do not give an exact definition of the employed three-body basis, rather we 
refer to its definition in \cite{DeE16}. Here we just mention
that we take as dynamical variables the two Jacobi vectors of the three-body system and
that the corresponding radial parts of the pair and the third-particle wave function are 
described by Laguerre polynomials times an exponential fall-off, accordingly we call 
this basis Jacobi basis. 

\begin{figure}
\centering
  \includegraphics[width=0.6\textwidth]{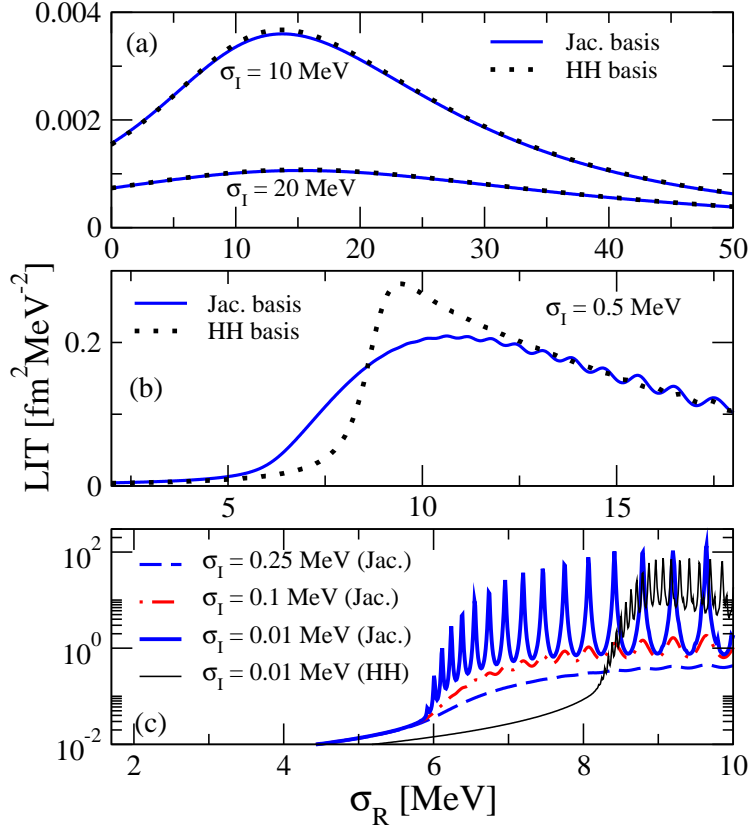}
\caption{LITs of  $R(E_\gamma)$ of Eq.~(\ref{response}) with HH and Jacobi basis
and various $\sigma_I$ as indicated in the figure}
\label{fig:1}       
\end{figure}

In Fig.~1 we show a comparison of LIT results obtained using HH and Jacobi basis systems.
For the larger $\sigma_I$ of 10 and 20 MeV there are only rather small differences
(Fig.~1a), but for the higher resolution of $\sigma_I = 0.5$ MeV one finds quite some 
difference in Fig.~1b. For the HH basis one sees a rather pronounced peak, whereas with 
the Jacobi basis the peak is broader and lower. The origin of this difference 
becomes clearer inspecting Fig.~1c. For the highest considered resolution 
($\sigma_I = 0.01$ MeV) one can actually identify the separate contributions 
of the various LIT states. One sees that with the Jacobi basis there is dense grid of LIT 
states starting right at the breakup threshold of about 5.8 MeV, whereas with the HH
basis there are no LIT states below the three-body breakup threshold at about 8 MeV
(note that with the MT-I/III potential one has binding energies of about 2.24 and 8.05 MeV
for $^2$H and $^3$He, respectively). Thus with the HH basis the strength below the
three-body breakup threshold is incorrectly shifted to energies above this threshold. 

The LIT results of Fig.~1c with the Jacobi basis and $\sigma_I = 0.01$ MeV
exhibit an increasing $\Delta E_n$ with 
growing energy, in other words a decreasing density of LIT states at higher energies. 
Fortunately, as Fig.~1c shows, one can work with rather high resolution 
($\sigma_I \approx 0.1$ MeV) close to the breakup threshold and then reduce the 
resolution step by step with increasing energy. As a rule of thumb one should impose 
$\sigma_I > \Delta E_n$. With this condition one avoids that a single LIT state is 
sticking out in the calculated transform thus leading to a sufficiently  
smooth LIT. In fact as described in \cite{DeE16} we use quite a number 
of different $\sigma_I$-values for the inversion of the LIT.

\begin{figure}
\centering
  \includegraphics[width=0.55\textwidth]{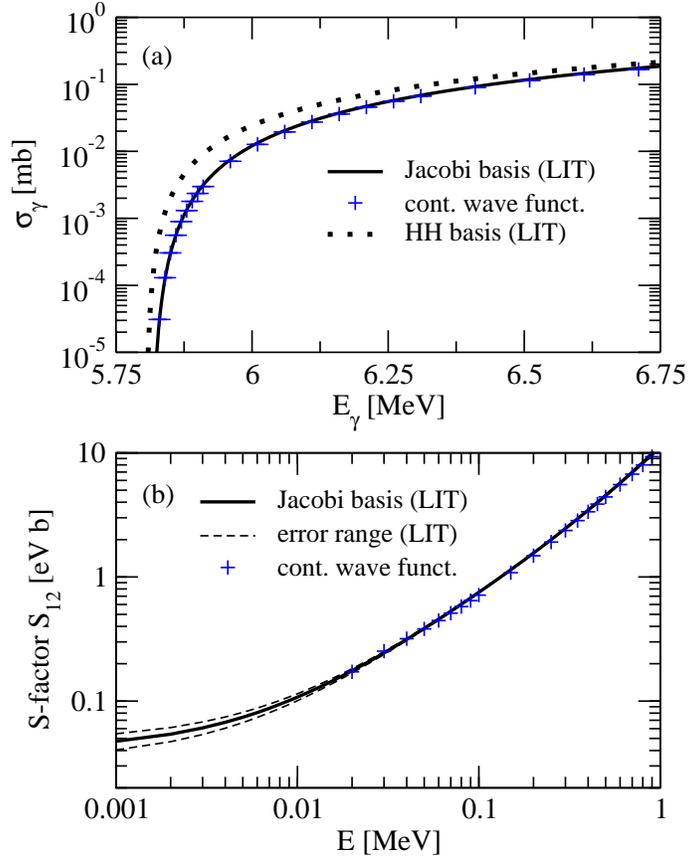}
\caption{$\sigma_\gamma$ (a) and $S_{12}$ (b) at low energies (see also text)}
\label{fig:2}       
\end{figure}

In Fig.~2a we show the results for the $\sigma_\gamma$ of $^3$He at
low energies. One sees that the inversions results from the LIT  
calculated with HH and Jacobi basis systems are rather different (note
the logarithmic scale for the cross section). On the other hand both results exhibit
an exponential threshold behaviour, however, this had to be expected, because the proper 
threshold behaviour is implemented in the applied inversion method \cite{EfL07}.
As already mentioned we also perform a direct calculation of $\sigma_\gamma$, where we 
have to compute the deuteron-proton continuum wave
functions. In Fig.~2a one observes an excellent agreement of these results with 
those from the inversion of the LIT with the Jacobi basis. 

Since the low-energy cross section is dominated by the exponential increase due to the 
Gamow factor it is interesting to see how the comparison of the results 
illustrated in  Fig.~2a looks like after a multiplication with the Gamow factor. As Eq.~(\ref{S-fac}) shows 
this brings us essentially to the $S$-factor $S_{12}$, which is shown in Fig.~2b. It is 
readily seen that the excellent agreement remains intact after the multiplication. 
In Fig.~2b we show in addition an error range due to the LIT inversion, 
it is given by the space between the two dashed curves (for an explanation how
the error is obtained see \cite{DeE16}). In principle one could reduce the error range 
by a calculation of the LIT with a larger Jacobi basis. 

Summarizing our work we may conclude that the LIT approach offers a viable alternative 
for the calculation of $S$-factors.   



\end{document}